\documentstyle[12pt]{article}

\catcode`@=12
\begin{document}

\thispagestyle{empty} 
\begin{flushright} UCRHEP-T237\\October 1998\
\end{flushright}
\vspace{0.5in}

\begin{center}
{\Large {\bf Alternative Interpretation of the Tevatron Top Events\\[0pt]
}} \vspace{1.0in} {\bf Darwin Chang, We-Fu Chang\\[0pt]
} \vspace{0.1in} {\sl NCTS and Department of Physics,\\[0pt]
} {\sl National Tsing-Hua University\\[0pt]
} {\sl Hsinchu 30043, Taiwan, Republic of China\\[0pt]
} \vspace{0.5in} {\bf Ernest Ma\\[0pt]
} \vspace{0.1in} {\sl Department of Physics, University of California\\[0pt]
} {\sl Riverside, California 92521, USA\\[0pt]
} \vspace{1.0in}
\end{center}

\begin{abstract}
We propose an alternative interpretation of the top events discovered 
at the Tevatron in 1995.  Given that the charge of the $b$ quark jet 
cannot be measured for the whole sample with certainty, the signal can be 
due to a quark of charge $-4/3$ at the reported mass, {\it i.e.} 174 GeV, 
while the top quark is actually heavier, say above 230 GeV.  We point out
in this paper that such a scenario is actually hinted at by the latest 
precision electroweak measurements of $Z$ decay.  To rule out this 
possibility in the future, the $b$ quark jet charge analysis has to become 
definitive, or the single "top" production cross section has to be
measured.
\end{abstract}

\newpage \baselineskip 24pt

It is generally believed that the top quark was discovered at Fermilab in 
1995.~\cite{1,2}  Its decay into a $W$ boson and a $b$ quark is an 
unmistakable signature.  However, unless the correlation between 
the charge of the $W$ boson and the 
charge of the $b$ quark jet can be measured with 
certainty,~\cite{3} the ``top quark" events may be 
either $W^+ b$ or $W^- b$.  If they are the former, then their identification 
as the decay product of $t$ from the expected $(t,b)_L$ doublet of the 
standard model is certainly justified.  If the latter, then an exotic quark 
of charge $-4/3$ is implied.  Of course, such a scenario appears to be 
totally unmotivated -- that is, until now. 
The updated precision measurements of electroweak parameters at the $e^+ e^-$ 
colliders LEP at CERN and SLC at SLAC reveal in fact an intriguing 
possibility that $m_t$ may 
be larger than about 230 GeV, and that the right-handed coupling of the
$b$ 
quark to the $Z$ boson may be significantly modified.

In this paper we analyze the 1998 precision electroweak data using the 
parameters $\epsilon_{1,2,3}$ and $\epsilon_b$,~\cite{4} but also allow 
$b_R$ to mix with an exotic quark $Q_1$ of charge $-1/3$ as part of the 
doublet $(Q_1,Q_4)_R$, where $Q_4$ has charge $-4/3$.~\cite{5,6}  We discuss 
how the current data~\cite{7} favor such an interpretation.  We then propose 
an exotic fourth family of quarks and leptons which is free of anomalies, 
together with a heavy Higgs scalar triplet~\cite{8} which supplies the 
neutrinos with Majorana masses.  We show that this model accounts for all the 
data, including the $Z \to b \bar b$ rate and forward-backward 
asymmetry.~\cite{9}  It has also an easily testable prediction in the 
single production of $Q_4$.

The phenomenological success of the standard gauge model of particle 
interactions is indisputable.  It is being tested experimentally at the 
one-loop level in terms of its calculable radiative corrections.  Consider the 
$\epsilon_{1,2,3}$ variables~\cite{10} which are purely weak radiative 
corrections to the two-point self-energy functions of the $W$ and $Z$ bosons. 
They are defined in such a way that they are zero in the standard-model tree 
approximation, keeping however the electromagnetic and strong-interaction 
radiative corrections.  Using as inputs the Fermi constant, $G_F$, the mass 
of the $Z$ boson, $m_Z$, and the electromagnetic fine-structure constant 
extrapolated to the $Z$ mass, $\alpha (m_Z)$, the $\epsilon_{1,2,3}$ 
variables can be determined from the experimental measurements of the 
partial width and forward-backward asymmetry of the decay of $Z$ to charged 
lepton pairs, and the mass of the $W$ boson, $m_W$.

Assuming lepton universality, the $Z \to l^- l^+$ partial width and 
forward-backward asymmetry are given by~\cite{4}
\begin{equation}
\Gamma_l = {G_F m_Z^3 \over 6 \pi \sqrt 2} \left( g_V^2 + g_A^2 \right) 
\left( 1 + {3 \alpha \over 4 \pi} \right),
\end{equation}
\begin{equation}
A_l^{FB} (\sqrt s = m_Z) = {3 g_V^2 g_A^2 \over (g_V^2 + g_A^2)^2},
\end{equation}
where
\begin{eqnarray}
g_A &=& - {1 \over 2} \left( 1 + {\epsilon_1 \over 2} \right), \\ 
g_V &=& - {1 \over 2} \left( 1 + {\epsilon_1 \over 2} \right) \left( 1 - 
4 \sin^2 \theta_{eff} \right).
\end{eqnarray}
In addition, $s_0^2$ (with $c_0^2 \equiv 1 - s_0^2$) is defined by
\begin{equation}
s_0^2 c_0^2 \equiv {\pi \alpha (m_Z) \over \sqrt 2 G_F m_Z^2},
\end{equation}
and $\sin^2 \theta_{eff}$ is related to $s_0^2$ by
\begin{equation}
{\sin^2 \theta_{eff} \over s_0^2} = 1 + {\epsilon_3 - c_0^2 \epsilon_1 \over 
c_0^2 - s_0^2}.
\end{equation}
With $\alpha (m_Z)^{-1} = 128.90$, it has been shown that~\cite{4}
\begin{equation}
\Gamma_l = 83.563~{\rm MeV}~(1 + 1.20 \epsilon_1 - 0.26 \epsilon_3),
\end{equation}
and
\begin{equation}
A_l^{FB} = 0.01696 ~(1 + 34.72 \epsilon_1 - 45.15 \epsilon_3).
\end{equation}
Given the latest experimental values~\cite{7}
\begin{equation}
\Gamma_l = 83.90 \pm 0.10~{\rm MeV},
\end{equation}
and
\begin{equation}
A_l^{FB} = 0.01683 \pm 0.00096,
\end{equation}
we find
\begin{eqnarray}
\epsilon_1 &=& (4.1 \pm 1.2) \times 10^{-3}, \\ 
\epsilon_3 &=& (3.3 \pm 1.8) \times 10^{-3}.
\end{eqnarray}
Using~\cite{4}
\begin{equation}
m_W^2 / m_Z^2 = 0.768905 ~(1 + 1.43 \epsilon_1 - 1.00 \epsilon_2 - 0.86 
\epsilon_3),
\end{equation}
and the latest experimental values~\cite{7}
\begin{equation}
m_W = 80.39 \pm 0.06 ~{\rm GeV}, ~~~ m_Z = 91.1867 \pm 0.0021 ~{\rm GeV},
\end{equation}
with Eqs.(11) and (12), we find
\begin{equation}
\epsilon_2 = (-7.8 \pm 2.8) \times 10^{-3}.
\end{equation}
The above values for $\epsilon_{1,2,3}$ agree very well with those obtained 
a year ago~\cite{4} based on earlier data.~\cite{11}  They are also very 
consistent with the Tevatron determination of $m_t = 173.8 \pm 5.0$ GeV. 
Clearly there is no discrepancy with the standard model as far as leptons 
are concerned.

Consider now $Z \to b \bar b$ decay.  There are 3 measured quantities: 
the partial width divided by the hadronic width, $R_b$, by both LEP and 
SLC, the forward-backward asymmetry at the $Z$ pole, $A_{FB}^{0,b}$, by 
LEP, and the left-right asymmetry, $A_b$, by SLC.  Theoretically, this 
process differs from all others by the important fact that the $t$ quark 
contributes to the vertex correction through the expected $(t,b)_L$ doublet, 
whereas $\epsilon_{1,2,3}$ contributes universally to all $Z$ decays. 
The effective left-handed and right-handed couplings of the $b$ quark to the 
$Z$ boson are given in the standard model by
\begin{eqnarray}
g_{bL} &=& \left( 1 + {\epsilon_1 \over 2} \right) \left[ - {1 \over 2} 
(1 + \epsilon_b) + {1 \over 3} \sin^2 \theta_{eff} \right], \\ 
g_{bR} &=& \left( 1 + {\epsilon_1 \over 2} \right) {1 \over 3} \sin^2 
\theta_{eff}.
\end{eqnarray}
Note that the $t$ quark contributes to only $g_{bL}$ through $\epsilon_b$.  
Several years ago when there was a large experimental $R_b$ excess, many 
theoretical attempts were made to increase the magnitude of $g_{bL}$ 
by postulating new contributions to $\epsilon_b$, whereas $g_{bR}$ was left 
untouched.  At that time, the measurement errors of $A_{FB}^{0,b}$ and 
$A_b$ were large enough so that the above approach was justified.  However, 
since about a year ago, there has been a small but theoretically very 
important shift in the data.  $R_b$ is now just one standard deviation 
above the theoretical prediction, but both $A_{FB}^{0,b}$ and $A_b$ are two 
standard deviations below.  Since the former is proportional to $g_{bL}^2 + 
g_{bR}^2$ and the latter are proportional to $g_{bL}^2 - g_{bR}^2$, this 
turns out to imply that~\cite{9} the magnitude of $g_{bL}$ is actually 
\underline {smaller} 
than the theoretical prediction and that of $g_{bR}$ is much greater. 
Hence an intriguing possibility exists that the shift of $g_{bL}$ is due 
to a much larger $m_t$ and that the shift of $g_{bR}$ is due to the mixing 
of $b_R$ with a heavy quark doublet $(Q_1,Q_4)_R$, where $Q_4$ has charge 
$-4/3$,~\cite {5,6} and its decay into $W^- b$ at the Tevatron 
was observed and assumed to be $W^- \bar b$ from $\bar t$.

>From the latest experimental values~\cite{7}
\begin{equation}
R_b = 0.21656 \pm 0.00074, ~~~ A_{FB}^{0,b} = 0.0991 \pm 0.0021, ~~~ 
A_b = 0.856 \pm 0.036,
\end{equation}
the couplings $g_{bL}$ and $g_{bR}$ have been determined:~\cite{9}
\begin{equation}
g_{bL} = -0.4159 \pm 0.0024, ~~~ g_{bR} = 0.1050 \pm 0.0090.
\end{equation}
Using $m_t = 174$ GeV, $m_H = 100$ GeV, and $\sin^2 \theta_{eff} = 0.23125 
\pm 0.00023$,~\cite{12} the standard model yields~\cite{9}
\begin{equation}
g_{bL}^{SM} = -0.4208, ~~~ g_{bR}^{SM} = 0.0774.
\end{equation}
>From Fig.~2 of Ref.~[9], it is seen that the standard-model point is just 
outside the 99\% confidence-level contour of the data.  From $g_{bL}$, 
we now find
\begin{equation}
\epsilon_b = (-15.7 \pm 4.9) \times 10^{-3}.
\end{equation}
This is a dramatically new result.  Previous analyses force a fit to 
\underline {both} $R_b$ and the asymmetries with $\epsilon_b$, but since 
$g_{bR}$ cannot be changed, the former tends to increase $g_{bL}$ and 
the latter tend to decrease it.  The end result is a much smaller magnitude 
for $\epsilon_b$ which is consistent with $m_t = 174$ GeV.  Now from Eq.(21), 
we obtain
\begin{equation}
m_t = 274 ^{+40}_{-47} ~{\rm GeV},
\end{equation}
where we have approximated $\epsilon_b$ by its leading contribution, 
$-G_F m_t^2 / 4 \pi^2 \sqrt 2$.

Next we change $g_{bR}$ by mixing $b$ ($I_{3R} = 0$) with a heavy 
quark $Q_1$ of $I_{3R} = 1/2$.~\cite{5,6}  We then have
\begin{eqnarray}
g_{bR} &=& \left( 1 + {\epsilon_1 \over 2} \right) \left[ {1 \over 3} \sin^2 
\theta_{eff} \cos^2 \theta_b + \left( {1 \over 2} + {1 \over 3} \sin^2 
\theta_{eff} \right) \sin^2 \theta_b \right] \nonumber \\ 
&=& \left( 1 + {\epsilon_1 \over 2} \right) \left( {1 \over 3} \sin^2 
\theta_{eff} + {1 \over 2} \sin^2 \theta_b \right).
\end{eqnarray}
Using Eq.(19), we find
\begin{equation}
\sin^2 \theta_b = 0.0554 \pm 0.0180.
\end{equation}
To be more specific, we propose an exotic fourth family of quarks and leptons. 
The $b_R$ singlet of the standard model becomes
\begin{equation}
b_R \cos \theta_b - Q_{1R} \sin \theta_b \sim (3,1,-1/3),
\end{equation}
where its $SU(3)_C \times SU(2) \times U(1)_Y$ representation content is also 
displayed. We then add
\begin{equation}
\left[ \begin{array}{c} Q_1 \cos \theta_b + b \sin \theta_b \\ Q_4 \end{array} 
\right]_R \sim \left( 3, 2, -{5 \over 6} \right),
\end{equation}
\begin{equation}
Q_{1L} \sim (3,1,-1/3), ~~~ Q_{4L} \sim (3,1,-4/3),
\end{equation}
\begin{equation}
\left[ \begin{array}{c} L_3 \\ L_2 \end{array} \right]_R \sim \left( 1,2,
{5 \over 2} \right), ~~~ \begin{array}{c} L_{3L} \sim (1,1,3), \\ L_{2L} \sim 
(1,1,2). \end{array}
\end{equation}
Anomalies cancel in the above because $Y_L = -3 Y_Q$ as in the standard model.
Within this model, Eq.(16) and Eq.(23) are valid to good accuracy with
$\epsilon_b$ again due to the top loop. 
In the following, we will take $m_4 = 174$ GeV for $Q_4$ to account for the 
Tevatron ``top'' events, and the true $m_t$ will be chosen to be higher, say 
about 230 GeV, to be consistent with Eq.(22).

We now calculate the extra contributions of the exotic fourth family of our 
model (relative to the standard model
with $m_t = 174$ GeV) to 
the variables $\epsilon_{1,2,3}$.~\cite{13} 
First,
\begin{equation}
\Delta \epsilon_3 = {\alpha \over 24 \pi s_0^2} \left( 3 + 5 \ln {m_1^2 
\over m_4^2} + 1 - 5 \ln {m_3^2 \over m_2^2} - \ln \left[ {m_t \over 174 
~{\rm GeV}} \right]^2 \right).
\end{equation}
For illustration, let $m_1 = 200$ GeV, $m_2 = 100$ GeV, $m_3 = 200$ GeV, 
then $\Delta \epsilon_3 = -0.93 \times 10^{-3}$, which is in fact a
better fit to the present data.  For heavier top mass, the fit improves
even more ({\it e.g.} for $m_t = 274$ GeV, $\Delta \epsilon_3 = -1.09 \times
10^{-3}$).
Note that the above values are chosen so that 
the processes $e^- e^+ \to L_2^{--} L_2^{++}$ and $Q_1 \bar b + b \bar Q_1$ 
are currently kinematically forbidden.  Second,
\begin{equation}
\Delta \epsilon_2 = - {\alpha \over 24 \pi s_0^2} \left( 3 g(m_1^2,m_4^2) 
+ g(m_3^2,m_2^2) + 3 \ln \left[ {m_t \over 174 ~{\rm GeV}} \right]^2 \right),
\end{equation}
where
\begin{equation}
g(x,y) = - {5 \over 3} + {4xy \over (x-y)^2} + {(x+y)(x^2-4xy+y^2) \over 
(x-y)^3} \ln {x \over y}.
\end{equation}
For the same sample values, we obtain $\Delta \epsilon_2 = -0.92 \times 
10^{-3}$, which is again a better fit to the present data.  Third,
\begin{equation}
\Delta \epsilon_1 = {\alpha \over 16 \pi s_0^2 c_0^2 m_Z^2} \left( 
3 f(m_1^2,m_4^2) + f(m_3^2,m_2^2) + 3 [ m_t^2 - (174 ~{\rm GeV})^2 ] 
\right),
\end{equation}
where
\begin{equation}
f(x,y) = x + y - {2xy \over x-y} \ln {x \over y}.
\end{equation}
For the same sample values, we obtain $\Delta \epsilon_1 = 8.7 \times 
10^{-3}$, which is of course a disaster.  
To fit present data, we need 
another source of $\epsilon_1$ which gives, say, $- 9 \times 10^{-3}$. 
Fortunately, there is already such a source in the form of a heavy Higgs 
scalar triplet~\cite{8}
\begin{equation}
\xi \equiv (\xi^{++}, \xi^+, \xi^0) \sim (1,3,1),
\end{equation}
which allows the standard-model neutrinos to acquire Majorana masses. 
After all, there is now a good deal of experimental evidence for neutrino 
oscillations and the minimal standard model should be extended to include 
nonzero neutrino masses.~\cite{14}  Let $\xi^0$ acquire a nonzero vacuum 
expectation value $u$ in addition to the standard-model Higgs doublet 
$\langle \phi^0 \rangle = v$, then in the tree approximation,
\begin{equation}
\left[ {G_F \over \sqrt 2} \right]_{CC} = {1 \over 4 v^2 + 8 u^2}, ~~~ 
\left[ {G_F \over \sqrt 2} \right]_{NC} = {1 \over 4 v^2 + 16 u^2},
\end{equation}
where $CC$ denotes charged current and $NC$ denotes neutral current.  The 
contribution to $\epsilon_1$ is thus $-2 u^2/(v^2 + 4 u^2)$ and for 
$u = 12$ GeV, the desired shift is obtained.  
This value of $u$ also 
implies that $m_\xi$ can be rather large.  Using the relationship~\cite{8} 
\begin{equation}
u \simeq - \mu v^2 / m_\xi^2,
\end{equation}
and setting the trilinear coupling $\mu$ of $\xi^\dagger \phi \phi$
equal to $m_\xi$, we obtain $m_\xi = 2.6$ TeV.  This in turn implies that 
the one-loop contributions of $\xi$ to $\epsilon_{1,2,3}$ should be 
negligible.  To check this, we have calculated the shifts of 
$\epsilon_{1,2,3}$ due to $\xi$.  Both $\epsilon_3$ and $\epsilon_2$ are 
corrected by $-\alpha/(12 \pi s_0^2)$ times a factor of order $v^2/m_\xi^2$. 
Hence these contributions are of order $10^{-5}$.  In the case of 
$\epsilon_1$, we find the leading contribution to be
\begin{equation}
\Delta \epsilon_1 = {\alpha \over 4 \pi s_0^2 c_0^2 m_Z^2} \left( m_{++}^2 
- 2 m_+^2 + m_0^2 \right),
\end{equation}
but this is zero because the sum rule $m_{++}^2 - m_+^2 = m_+^2 - m_0^2$ 
is valid~\cite{15} in the approximation that the mass differenecs in $\xi$
are only due to scalar doublets. We estimate 
the next term to be at most of order $10^{-5}$ to $10^{-4}$.

\underline {Crucial tests of the $Q_4$ hypothesis}.  The unambiguous method 
is to distinguish a $b$ jet from a $\bar b$ jet by its charge, if it can be 
done with certainty.~\cite{3}  A potentially easier way is to measure the 
single "top" production cross section~\cite{16} in the future Run II of
the 
Tevatron scheduled to begin in 2000.  Comparing it against the $t \bar t$ 
cross section would determine $|V_{tb}|^2$.  If it turns out to be given by 
Eq.~(24) and not unity as expected in the standard model, it would be an 
indication that the top events may be due to $Q_4$.  On the other hand, 
it may accidentally be so because there is a heavier but ordinary fourth 
family of quarks, {\it i.e.} $(t',b')_L$, $t'_R$, and $b'_R$.  In any case, 
if the single ``top" production cross section is actually only a few
percent of its expected value, the number of events may be too small to be
observed above background.

In conclusion, we point out in this paper that the top events may be due to 
a heavy quark of charge $-4/3$ instead of $2/3$.  This would come about if 
$m_t$ is actually much larger than 174 GeV, and $b_L$ or $b_R$ mixes with a 
heavy quark $Q_1$ of charge $-1/3$ whose doublet partner $Q_4$ has charge 
$-4/3$.  In particular, the case with $b_R$ mixing with $(Q_1,Q_4)_R$ can
result in a better fit of the present data on $R_b$, $A_{FB}^{0,b}$, and
$A_b$ than the standard model.  The same fit also favors $m_t > 230$ GeV.
As discussed in the previous paragraph, 
this hypohesis will be tested in the future Run II of the Tevatron.
\vspace{0.3in}
\begin{center} {ACKNOWLEDGEMENT}
\end{center}

We thank C.-Q. Geng for discussion of anomaly, and G.-P. Yeh, M.-J. Wang,
and G. Velev for discussions on CDF physics.  
EM is supported in part by the U.~S.~Department of Energy under Grant No.
DE-FG03-94ER40837; DC and WFC are supported by a grant from the
National Science Council of ROC.

\newpage

\begin{center} {APPENDIX}
\end{center}

Our model of an exotic fourth family naturally has a softly broken 
discrete $Z_2$ symmetry, under which all the exotic fermions are odd
and the ordinary 
ones are even.  It is broken only by the explicit soft term $m' \bar Q_{1L} 
b_R$.  Hence the $2 \times 2$ mass matrix linking ($\bar b_L, \bar Q_{1L}$) 
with ($b_R, Q_{1R}$) is given by
\begin{equation}
{\cal M}_{b Q_1} = \left[ \begin{array}{c@{\quad}c} m_b & 0 \\ m' & m_1 
\end{array} \right].
\end{equation}
Thus $\sin \theta_b \simeq m'/m_1$ and $m'$ can be small naturally because
it breaks this discrete $Z_2$ symmetry.


\begin{thebibliography}{99}
\bibitem{12} With this value of $\sin^2 \theta_{eff}$ from leptonic data 
alone, the LEP and SLC measurements~\cite{7} are only +0.85$\sigma$ and 
$-0.77\sigma$ away, instead of $+0.54\sigma$ and $-1.78\sigma$ if the 
$b \bar b$ data are included.
\end{thebibliography}
\end{document}